\documentclass[reqno]{amsart}

\usepackage[a4paper,margin=1.08in]{geometry}
\usepackage[T1]{fontenc}
\usepackage{lmodern}
\usepackage{microtype}
\usepackage{setspace}
\usepackage{xcolor}
\usepackage{amsmath,amssymb,amsthm,mathtools,mathrsfs}
\usepackage{bm}
\usepackage{enumitem}
\numberwithin{equation}{section}

\usepackage[numbers,sort&compress]{natbib}
\usepackage[colorlinks=true,
            linkcolor=blue!55!black,
            citecolor=blue!55!black,
            urlcolor=blue!55!black]{hyperref}

\newtheorem{theorem}{Theorem}[section]

\newtheorem{lemma}[theorem]{Lemma}

\theoremstyle{definition}
\newtheorem{definition}[theorem]{Definition}
\theoremstyle{plain}
\newtheorem{remark}[theorem]{Remark}
\theoremstyle{plain}
\newtheorem*{maintheorem*}{Main Theorem}
\newtheorem*{corollary*}{Corollary}

\newcommand{\M}{\mathcal{M}}
\newcommand{\D}{\mathcal{D}}
\newcommand{\A}{\mathcal{A}}
\newcommand{\Hh}{\mathcal{H}^{+}}

\newcommand{\dd}{\,\mathrm{d}}
\newcommand{\T}{\boldsymbol{T}}
\newcommand{\K}{\boldsymbol{K}}
\newcommand{\Ss}{\mathscr{S}}
\newcommand{\J}{\boldsymbol{J}}
\newcommand{\Ll}{\boldsymbol{L}}
\newcommand{\lie}{\mathcal{L}}
\newcommand{\tr}{\mathrm{tr}}
\newcommand{\ddiv}{\mathrm{div}}



\begin{document}
\title[Nonexistence of Multiple Black Holes]{ 
Nonexistence of Multiple Black Holes \\
in Stationary Analytic Vacuum Spacetimes}

\author{YISHEN HUANG}
\author{JUNBIN LI}
\author{XI-PING ZHU}

\thanks{All authors are supported by National Key R\&D Program of China (No. 2022YFA1005400). J. Li is also supported by NSFC (12326602, 12141106). }

    \begin{abstract}

We prove the nonexistence of non-rotating and degenerate horizon components in the stationary, asymptotically flat, $I^+$-regular, analytic vacuum spacetime. This result rules out the final case in the classification of stationary analytic black-hole spacetimes. Consequently, the stationary, asymptotically flat, \(I^+\)-regular, analytic vacuum black hole uniqueness conjecture is settled affirmatively in full.

\end{abstract}
  
\maketitle

    \section{Introduction}

\label{introduction}

The rigidity or uniqueness problem is one of the long-standing issues in general relativity. It is anticipated that the metric of a stationary spacetime is of a simple form.
For spacetimes without black holes, several rigidity results have been obtained. 
Lichnerowicz \cite{Lichnerowicz1955theories} proved that an asymptotically flat vacuum spacetime admitting a globally defined static Killing field must be Minkowski spacetime.
Anderson \cite{Anderson2000} proved that if a vacuum spacetime is geodesically complete, chronological (which means that the spacetime contains no closed timelike curves), and admits a globally timelike complete Killing field, then it is flat,
whereas Chen removed the chronological condition in \cite{ChenBingLong}.

For spacetimes with black holes, it comes to a long-standing problem, which is named as the black hole uniqueness conjecture. This conjecture asks whether every \textit{regular}, stationary vacuum black-hole spacetime must be the Kerr spacetime. 

In the setting of \textit{analytic}\footnote{Throughout this paper, "analytic" always means real analytic.} spacetimes,
this conjecture is settled when the spacetime has only one black hole, i.e., the horizon is connected. The first non-trivial result concerning this question is due to Hawking, known as Hawking's rigidity theorem \cite{Hawking-Ellis}. He proved that a stationary analytic spacetime must admit an axisymmetric Killing field under the condition that the interior of the ergoregion intersects the horizon (this condition is also phrased by saying that the horizon is rotating, see \cite{Chrusciel2008} or the definition of the term "rotating horizon" in the following). 
We stress that Hawking's rigidity theorem also applies to multi-black-hole case, since its argument is local on the horizon.

Therefore, a stationary analytic spacetime with a rotating horizon is also an axisymmetric spacetime, and the vacuum Einstein equations can then reduce to a singular harmonic map. Carter \cite{Carter1973} and Robinson \cite{Robinson1975Uniqueness} proved that the solution to this harmonic map is unique when the horizon is connected and non-degenerate. 
When the horizon is connected and degenerate, the uniqueness of solutions to this harmonic map is proved by Meinel--Ansorg--Kleinw\"achter--Neugebauer--Petroff \cite{Meinel-Reinhard-Ansorg-Marcus-Kleinwachter}, Amsel--Horowitz--Marolf--Roberts \cite{Amsel_2010}, Figueras--Lucietti \cite{Lucietti-Figueras}, and Chru\'sciel--Nguyen \cite{Chrusciel-Nguyen}. 
The proofs for the above two cases have assumed the existence of the bifurcate surface, or the cross-section of the horizon, and the simply connectedness of the domain of outer communication. Thus, a suitable framework is provided under the condition of $I^+$-regularity (see the definition in \cite{Chrusciel2008} or in the following), which was first introduced by Chru\'sciel and Costa in \cite{Chrusciel2008}. 
We emphasize that in the above two cases, the analyticity of the metric is only used in Hawking's rigidity theorem to obtain a globally axisymmetric Killing field.
Therefore, a stationary, asymptotically flat, $I^+$-regular, analytic vacuum spacetime with a rotating connected horizon must be Kerr spacetime. 

For a stationary vacuum spacetime with a connected, non-rotating and non-degenerate horizon, one can also prove the black hole uniqueness conjecture. Indeed, such a spacetime is static, which was first proved by Sudarsky and Wald in \cite{Sudarsky-Wald1992}, and a simplified proof is given by Chru\'sciel and Costa in \cite{Chrusciel2008}. 
Roughly speaking, this proceeds in two steps: first, extend the domain of outer communication to find out the bifurcate surface \cite{Wald-Racz1996}; second, construct a maximal hypersurface, whose interior boundary is that bifurcate surface, and which can extend to the spatial infinity \cite{Chrusciel-Wald_1994}. Finally, by integrating the constraint equations on that maximal hypersurface with that boundary condition in the first step, the second fundamental form is forced to vanish everywhere, so that the domain of outer communication is static \cite{Chrusciel2008}. It has been proved by Israel \cite{Israel1967} that a static asymptotically flat vacuum spacetime with a connected and non-degenerate horizon must be Schwarszchild spacetime, which is the first major result for the black hole uniqueness conjecture. 
We stress that the above argument does not require analyticity of the metric. 
Therefore, a stationary, asymptotically flat, $I^+$-regular, smooth vacuum spacetime with a connected, non-rotating and non-degenerate horizon must be Schwarszchild spacetime, which also belongs to the Kerr solutions.

The case that the connected horizon is non-rotating and degenerate can be readily ruled out by the Komar identity for the ADM mass and the positive mass theorem (see, e.g., \cite{Chrusciel1999}). We stress that this case does not require analyticity of the metric.
Therefore, a stationary, asymptotically flat, $I^+$-regular, smooth vacuum spacetime with a non-rotating connected horizon must be Schwarszchild spacetime. 

As a consequence, the black hole uniqueness conjecture for a single black hole in an analytic spacetime is settled. More precisely, since the region of observational interest is the domain of outer communication, the uniqueness referred to above is in fact the uniqueness of the domain of outer communication.
We state the above results together rigorously in the following theorem.
\begin{theorem}
    \label{thm:uniq-single-bh}
    The domain of outer communication of a stationary, asymptotically flat, $I^+$-regular, analytic vacuum spacetime with a connected horizon, is isometric to that one of the Kerr spacetime.
\end{theorem}

To discuss multi-black-hole case, let us first introduce some terminology. Suppose $(\M,g)$ is a four-dimensional spacetime
\footnote{Throughout this paper, we only consider four-dimensional spacetimes.}
. We say it is an analytic spacetime, if there is an analytic atlas on $\M$ such that the components of the metric $g$ are analytic in each coordinate chart. A Killing vector field $\K$ is said to be complete if its orbits are complete. Furthermore, $(\M,g)$ is said to be a stationary, asymptotically flat, $I^+$-regular, analytic vacuum spacetime, if it is an analytic spacetime, satisfies the vacuum Einstein equation $R_{\mu \nu} = 0$, and obeys the following hypotheses:
\begin{enumerate}[label=\textup{(H\arabic*)},leftmargin=2.8em]
    \item (Asymptotic flatness). 
        \label{H1:AF}
        $\M$ contains a spacelike hypersurface $\Ss_{ext}$ diffeomorphic to $\mathbb{R}^3 \setminus B(R)$, where $B(R)$ is an open coordinate ball with radius $R$, and on $\Ss_{ext}$ there is a coordinate system $\{ x^i,\ i=1,2,3\}$ such that the metric $h$ induced by $g$ and the second fundamental form $k$ of $\Ss_{ext}$ in $\M$ take the asymptotic forms:
        \begin{align*}
            &h_{ij} = \left( 1 + \frac{2M}{r} \right) \delta_{ij} + o_2 (r^{-1})\ ,\\
            &k_{ij} = o_2 (r^{-1})\ ,
        \end{align*}
        where $r = \sqrt{\sum_{i=1}^{3}(x^i)^2}$, and a function is $o_l (r^{-p})$ if it is $o(r^{-p})$ and its m-th derivative is $o(r^{-p-m})$ for any $0 \le m \le l$. The constant $M$ is the ADM mass of the spacetime. $\Ss_{ext}$ is also said to be an asymptotically flat end.
    \item (Stationary Killing field).
    \label{H2:stationary}
        $(\M,g)$ is a stationary spacetime, in the sense that there is a Killing field $\T \in \mathcal{T}(\M)$ which is timelike in the asymptotic infinite region $\displaystyle \M_{ext} \overset{\triangle}{=} \bigcup_{t \in \mathbb{R}} \Phi_t (\Ss_{ext})$, where $\{\Phi_t \}$ is the one-parameter isometric group generated by $\T$. $\T$ is also said to be a stationary Killing field.
    \item ($I^+$-regularity).
        \label{H3:I+regular}
        Let $(\M,g)$ be a spacetime with an asymptotically flat end $\Ss_{ext}$ and a stationary Killing field $\T$.
        $(\M,g,\T)$ is said to be an $I^{+}$-regular spacetime, if $\T$ is a complete Killing field, if the domain of outer communication $\D \overset{\triangle}{=} I^{-}(\M_{ext}) \cap I^{+}(\M_{ext})$ is globally hyperbolic, and if there is a spacelike, connected, acausal  hypersurface $\Ss$ in $\D$, consisting of the union of a compact set and finite number of asymptotic ends, such that $\Ss \supset \Ss_{ext}$, the closure $\bar{\Ss}$ of which is a topological manifold with boundary, and the boundary $\partial \bar{\Ss} = \bar{\Ss} \setminus \Ss$ is a topological manifold satisfying
    \begin{equation*}
        \partial \bar{\Ss} \subset \Hh \cap I^{+}(\M_{ext})\ ,
    \end{equation*}
    with $\partial \bar{\Ss}$ meeting every generators of the (future) horizon $\Hh \overset{\triangle}{=} \partial I^{-}(\M_{ext})$ precisely once.
\end{enumerate}

We also collect some concepts which are widely used in general relativity.

\begin{definition}[Black holes on an instant time slice]
      \label{def:BH}
      Denote the black-hole region by $\boldsymbol{B} = \M \setminus I^{-}(\M_{ext})$. For every spacelike hypersurface $\Sigma$,
      each connected component of the intersection $ \Sigma \cap \boldsymbol{B}$ is said to be a black hole on the instant $\Sigma$, and each connected component of the cross-section $\Sigma \cap \Hh$ is said to be the surface of a black hole. 
\end{definition}

\begin{definition}[Rotating and non-rotating black holes]
    \label{def:rotating and non-rootating}
    Let $(\M,g)$ be a stationary spacetime, and $\T$ be the stationary Killing field. If $\T$ is tangent to the null generators of a connected component $\Hh_i$ of the horizon $\Hh$ everywhere, we say the black hole $\Sigma \cap \Hh_i$ (or the horizon component $\Hh_i$) is non-rotating, otherwise is rotating.
\end{definition}

Since the spacetime we consider is at least smooth (or analytic) and $I^+$-regular, the horizon $\Hh$ is smooth (and is analytic if $g$ is analytic, see Section 4.3 of \cite{Chrusciel2008}), we can talk about the covariant derivative of the null normal vector to the horizon, and thus the surface gravity is well-defined, and we can talk about the degeneracy of the horizon.

\begin{definition}[Surface gravity]
    \label{def:surface gravity}
    The surface gravity $\kappa$ of a horizon component $\Hh_i$, is defined by the equation:
    \begin{equation*}
        \nabla_l l = \kappa l\ ,
    \end{equation*}
    where $l$ is the tangent vector to the null generators of the (smooth) horizon component $\Hh_i$. 
    It can be proved that $\kappa$ is a constant everywhere on $\Hh_i$ (see, e.g., \cite{Wald-GR1984}).
    If $\kappa = 0$ on $\Hh_i$, we say the black hole $\Sigma \cap \Hh_i$ (or the horizon component $\Hh_i$) is degenerate, otherwise is non-degenerate. 
\end{definition}

\begin{definition}[Stationary and axisymmetric spacetimes]
    \label{def:stationary and axisymmetric}
    $(\M,g)$ is said to be a stationary and axisymmetric spacetime, in the sense that there are two  smooth and complete Killing fields $\T = \partial_t$ and $\K = \partial_{\phi}$, such that
    $(\M,g,\T)$ is a stationary spacetime; $\K$ is an axisymmetric Killing field which means that it has closed period orbits and generates a subgroup $SO(2)$ of the isometric group of the domain of outer communication 
    $\D$ of the spacetime $(\M,g)$. In addition, $\K$ is required to commute with $\T$, and the normal bundle of $\K$ and $\T$ is integrable, which means that they satisfy the Frobenius condition:
    \begin{align*}
        \T^{\flat} \wedge \K^{\flat} \wedge \dd \T^{\flat} = 0\ ,\quad
        \T^{\flat} \wedge \K^{\flat} \wedge \dd \K^{\flat} = 0\ ,
    \end{align*}
    where $\T^{\flat}$, $ \K^{\flat}$ are the dual one-forms of $\T$ and $\K$.
\end{definition}

When the stationary spacetime has multiple black holes, i.e., the horizon has more than one connected components,
it has been a long-standing conjecture that this case cannot occur. Intuitively speaking, this is because gravitational interaction gives rise to an attractive effect between the black holes, which would violate stationarity. 

Mathematically, if the stationarity condition is strengthened to staticity, then 
for the general static non-degenerate multi-black-hole case, it was shown by Bunting and Masood-ul-Alam \cite{Bunting-Masood} that such a spacetime must be the Schwarzschild spacetime, and thus the horizon has only one component. Chru\'sciel--Reall--Tod \cite{Chrusciel-degenerate} proved that degenerate black holes cannot exist in a static vacuum spacetime, and therefore the static vacuum spacetime is uniquely determined to be the Schwarzschild spacetime.

For the general case, from previous work (see, e.g., \cite{Chrusciel2008}), stationary, asymptotically flat, $I^+$-regular, analytic vacuum spacetimes fall into two classes. The first class is that at least one of the horizon components is rotating. As stated before, by Hawking's rigidity theorem, this spacetime is in fact not only stationary but also axisymmetric. Hence by the combination of the recent work by Han--Khuri--Weinstein--Xiong \cite{HKWX1} and an area-angular momentum relation by Hennig--Ansorg--Cederbaum \cite{Hennig-Cederbaum-Ansorg1}, the horizon is indeed connected, and thus the spacetime is the Kerr spacetime.

Rigorously speaking, suppose that $ \Hh = \bigcup_{i=1}^{N} \Hh_i$, where $\Hh_i\ ,\ i=1,2,\dots,N$ are the connected components of the horizon $\Hh$. Then the main theorem 1.1 in \cite{HKWX1} states that
\begin{theorem}[Han--Khuri--Weinstein--Xiong, {\cite{HKWX1}}]
    \label{thm:HKWX1}
    Let $N>0$ be the number of the black holes in a stationary, axisymmetric, asymptotically flat, smooth vacuum spacetime $(\M,g)$. Denote by $\mathcal{J}_i$ the angular momentum of the i-th black hole. Assume that 
    \begin{equation*}
        \mathcal{J}_i \neq 0
    \end{equation*}
    for every degenerate black hole; no hypothesis is imposed on the angular momenta of non-degenerate black holes. Then $N=1$ and the spacetime $(\M,g)$ is the Kerr spacetime. 
\end{theorem}

We note that for any degenerate black hole, there is in fact an area-angular momentum relation proved by Hennig--Ansorg--Cederbaum \cite{Hennig-Cederbaum-Ansorg1}, which states that
\begin{lemma}[Hennig--Ansorg--Cederbaum, {\cite{Hennig-Cederbaum-Ansorg1}}]
    \label{lem:HCA}
    Let $(\M,g)$ be a stationary, axisymmetric, \textit{regular}, smooth vacuum spacetime, with possibly multiple black holes. Suppose that in $(\M,g)$ the area of a degenerate black hole is $A_i$, and its angular momentum is $\mathcal{J}_i$, then
    \begin{equation*}
        8 \pi |\mathcal{J}_i| = \A_i\ .
    \end{equation*}
\end{lemma}

Here the word "\textit{regular}" means that the components of the metric $g$ is regular under the horizon-penetrating coordinates, or equivalently, the horizon component is assumed to have a smooth cross-section, so that the area of a degenerate black hole is well-defined and positive. This property is ensured by the $I^+$-regularity. As a consequence, the angular momentum of a degenerate black hole in the stationary, axisymmetric, asymptotically flat, $I^+$-regular, smooth vacuum spacetime, cannot be zero. The assumption for the angular momentum of degenerate black holes in Theorem \ref{thm:HKWX1} can therefore be eliminated. We thus obtain the uniqueness result for the first class of stationary, asymptotically flat, $I^+$-regular, analytic vacuum spacetimes with multiple black holes.

The second class is that all of the horizon components are non-rotating. Again, as stated before, by the Komar identity for the ADM mass, at least one component is non-degenerate. If, in addition, all components are non-degenerate, then one can still find a maximal hypersurface, and by \cite{Chrusciel2008} the spacetime is in fact static, and thus by the classical static uniqueness theorem by Bunting and Masood-ul-Alam \cite{Bunting-Masood}, the spacetime is in fact the Schwarszchild spacetime.

The final case in the second class stated above is that at least one horizon component is degenerate. This type of horizon components was expected not to exist by Chru\'sciel--Nguyen \cite{Chrusciel-Nguyen}. Our main theorem is to exclude this case.
\begin{maintheorem*}[Nonexistence of non-rotating and degenerate black holes]
\label{thm:nonexsitence of nonrotating and degenerate black holes}
    Let $(\M,g)$ be a stationary, asymptotically flat, $I^+$-regular, analytic vacuum spacetime, and $\Hh$ be the future horizon which may be disconnected. Then none of the connected components of $\Hh$ could be non-rotating and degenerate.
\end{maintheorem*}

We stress that since our argument only concerns a small neighbourhood of a non-rotating and degenerate horizon component, our main theorem can therefore apply to spacetimes containing both rotating and non-rotating horizon components.

Now combining all the foregoing results, we complete the classification of stationary analytic vacuum spacetimes, and obtain an affirmative answer to the black hole uniqueness conjecture in analytic spacetimes.
\begin{corollary*}[Uniqueness of stationary, asymptotically flat, analytic vacuum spacetimes]
\label{thm:Cw uniqueness}
    The domain of outer communication of a stationary, asymptotically flat, $I^+$-regular, analytic vacuum black-hole spacetime is isometric to that one of the Kerr spacetime. No hypothesis is imposed on the connectedness of the horizon.
\end{corollary*}

\begin{remark}
\label{rmk:electrovacuum1}
    In the stationary, axisymmetric and asymptotically flat electrovacuum spacetime, the authors of \cite{HKWX1} also have proven the uniqueness of Kerr-Newman spacetime in \cite{HKWX2}. In their paper, they also make an assumption to the possibly existing degenerate black holes, which says that for every degenerate black hole, either the angular momentum $\mathcal{J}_i \neq 0$ or the charge $Q_i \neq 0$. We also point out that there is an angular momentum-area-charge relation which has been proven in \cite{Hennig-Cederbaum-Ansorg2} by Hennig-Cederbaum-Ansorg, that for every degenerate black hole in \textit{regular}, stationary, axisymmetric, electrovacuum spacetime,
    \begin{equation*}
        (8 \pi \mathcal{J}_i)^2 + (4 \pi Q_i^2)^2 = \A_i^2\ .
    \end{equation*}
    Here the word "\textit{regular}" also means that the components of the metric $g$ is smooth and regular under the horizon-penetrating coordinates, or equivalently the horizon component has a smooth cross-section, 
    which are ensured by $I^+$-regularity.
    Therefore, the assumption to the possibly existing degenerate black holes in \cite{HKWX2} can be also eliminated by the $I^+$-regularity.
\end{remark}

We now outline the the proof of the main theorem. The entire argument is based on contradiction. Assume, for contradiction, there exists a non-rotating and degenerate connected component of the horizon. Then by the vacuum Einstein equations and the criterion for the first variation of the outer expansion of a degenerate horizon, we can find an axisymmetric Killing field $\J$ on the horizon, which was first obtained by Donajski--Lucietti in \cite{Lucietti_2023}. Based on their work, we calculate and prove that the Lie derivative of all the transversal derivatives of the metric with respect to $\J$ is indeed zero.
To show this, we use the vacuum Einstein equations restricted on the horizon component, rather than the NP formulae in \cite{Hawking-Ellis}. 

It is similar to the work by Hollands--Ishibashi \cite{Hollands-Ishibashi} when the horizon is \textit{rotating} and degenerate, and Hollands--Ishibashi--Wald \cite{Hollands-Ishibashi-Wald2007} when the horizon is \textit{rotating} and non-degenerate. 
But the key difference is that since the horizon they consider is rotating, those restricted Einstein equations will provide an extra term involving  derivatives of one order higher with respect to the candidate Killing field which is tangent to the null generators of the horizon. This extra term substantially simplifies the induction, so that their work focuses on higher dimensional situation, in which the spacelike Killing field constructed by the difference between the stationary Killing field and the candidate Killing field may fail to have closed orbits, and therefore they introduced a diophantine condition to avoid that issue.

However, in the case where the horizon component is \textit{non-rotating}, that extra term vanishes, and the axisymmetric Killing field cannot be constructed by the difference between the stationary Killing field and a vector field tangent to the null generators of the horizon component.
Fortunately, the variations of the vacuum Einstein equations restricted on the horizon component become elliptic equations for the transversal derivatives of the induced metric. By an energy estimate of the Lie derivative of those transversal derivatives with respect to $\J$ with an appropriate weight function, we can still prove that such a Lie derivative vanishes (see \textbf{Step 4} in Subsection \ref{subsec:3.2}).
Therefore by the simply connectedness of the domain of outer communication $\D$ and the analyticity of the metric $g$, $\J$ can be extended to be an axisymmetric Killing field all over $\D$ \cite{Nomizu1960}. 
Thus with direct verification, the spacetime becomes a stationary and axisymmetric vacuum spacetime in the sense of Definition \ref{def:stationary and axisymmetric}. By the combination of Theorem \ref{thm:HKWX1} and Lemma \ref{lem:HCA} as stated before, we conclude that the spacetime has only one black hole. Then one can use Theorem \ref{thm:uniq-single-bh} to obtain that the spacetime is actually a Kerr spacetime. But no member of the Kerr spacetimes has a non-rotating and degenerate horizon component, which leads to a contradiction.

The paper is organized as follows. In Section \ref{sec:preliminaries}, we set up our model and write out the vacuum Einstein equations restricted to the non-rotating and degenerate horizon under the Gaussian null coordinates. In Section \ref{sec: Axisymmetric Killing field}, for convenience and completeness, we repeat the argument by Dunajski--Lucietti in \cite{Lucietti_2023} to obtain an axisymmetric Killing field on the non-rotating and degenerate horizon component in Subsection \ref{subsec:3.1}. Then we extend it to the domain of outer communication in Subsection \ref{subsec:3.2}. In Section \ref{sec:The proof of the main theorem} we verify that the spacetime indeed becomes stationary and axisymmetric, which yields a result contradicting Theorem \ref{thm:uniq-single-bh} for the single-black-hole case, and obtain the nonexistence of non-rotating and degenerate horizon components.

\vspace{1em}
\noindent\textbf{AI Disclosure.} AI was used to verify the calculations in this paper. The idea on how to simplify the integral $II$ in \eqref{lie-Rab4} was suggested by AI.
    \section{Preliminaries}
\label{sec:preliminaries}

Consider the spacetime $(\M,g)$ to be stationary, smooth and $I^+$-regular.
Suppose $\Hh_i$ is a non-rotating and degenerate connected component of the horizon $\Hh$. By Theorem 4.5 of \cite{Chrusciel2008}, we can find a smooth two-surface $S= \Sigma \cap \Hh_i$, and we can impose the Gaussian null coordinates$\{ v,r,x^a \ | \ a=1,2 \}$ in a neighbourhood $\mathcal{U}$ of the non-rotating component $\Hh_i$, such that $\Hh_i = \{ r=0 \}$, $S = \{ r=0,v=0 \}$, and the spacetime metric $g$ takes the form
\begin{equation*}
    g = r \varphi \dd v^2 + 2 \dd v \dd r + 2rh_a \dd x^a \dd v + \gamma_{ab} \dd x^a \dd x^b\ ,
\end{equation*}
where $a,b = 1,2$, $\T = \partial_v$. Thus $\varphi=\varphi(r,x), h_a = h_a (r,x), \gamma_{ab}=\gamma_{ab}(r,x)$, where $x=(x^1,x^2)$, and $\gamma_{ab}(0,x)$ is the induced metric on $S$. 

Under this coordinates, the dual one-form of $\T$ is 
\begin{equation}
    \T^\flat = g(\T,\cdot) = r\varphi \dd v + \dd r + r h_a \dd x^a\ ,
\end{equation}
and $\T^\flat |_{r=0}= dr$.
By the degeneracy of $\Hh_i$, we have
\begin{equation*}
    0 = \kappa = -\frac{1}{2} \partial_r (r \varphi) \bigg|_{r=0}\ ,
\end{equation*}
and therefore $\varphi|_{r=0} = 0$, so we can assume that $\varphi= r F$, where $F = F(r,x^a)$ is a smooth function. Thus the metric in a neighbourhood of the non-rotating and degenerate component $\Hh_i$ takes the form
\begin{equation}
    g = r^2 F \dd v^2 + 2 \dd v \dd r + 2rh_a \dd x^a \dd v + \gamma_{ab} \dd x^a \dd x^b\ .
\end{equation}

The vacuum Einstein equations now can be written as the equations of the components under this coordinates:
\begin{align}
    0 = R_{rr} 
    &= -\frac{1}{2} \gamma^{ab}\gamma_{ab}'' + \frac{1}{4} \gamma^{cd}\gamma^{bd} \gamma_{ab}' \gamma_{cd}'\ ,\label{Rrr}\\
    0 = R_{ra} 
      &= \frac{1}{2 \sqrt{\det \gamma}} \partial_r \left[ \sqrt{\det \gamma}( (rh_a)' - rh^b \gamma_{ab}') \right] + \frac{1}{2} D^b \gamma_{ab}' - \frac{1}{2} D_a ( \gamma^{bc} \gamma_{bc}' ) \notag\\
      &= \left( \frac{1}{4} \gamma^{cd}\gamma_{cd}' h_a + \frac{1}{2} D^b \gamma_{ab}' - \frac{1}{2 } D_a (\gamma^{bc}\gamma_{bc}') + h_a' - \frac{1}{2} h^b \gamma_{ab}' \right) \notag\\
      &\quad\  + r \left( \frac{1}{4} \gamma^{cd}\gamma_{cd}' {h^e}' \gamma_{ae} + \frac{1}{2} {h_a}'' - \frac{1}{2} {h^b}' \gamma_{ab}' - \frac{1}{2} h^b \gamma_{ab}'' \right)\ , \label{Rra}\\
    0 = R_{ab}
      &= 
      Ric(\gamma)_{ab} + \frac{1}{2 \sqrt{\det \gamma}} \partial_r \left[
      \sqrt{\det \gamma} (2r D_{(a} h_{b)} + r^2 ( F-|h|^2 ) \gamma_{ab}'
      \right] 
      \notag \\
      &\qquad\qquad  
      + \frac{r}{2} D_c (h^c \gamma_{ab}') - \frac{1}{2} \left[ (rh_a)' - rh^c \gamma_{ac}' \right] \left[ (rh_b)' - rh^d \gamma_{bd}' \right] \notag\\
      &\qquad\qquad 
      - r\gamma_{c(a}' D^c h_{b)} - \frac{r^2}{2} (F - |h|^2) \gamma_{ac}' \gamma_{bd}' \gamma^{cd}
      \notag\\
      &= 
      \bigg( Ric(\gamma)_{ab} + D_{(a}h_{b)} - \frac{1}{2}h_a h_b \bigg) 
      \notag\\
      &\quad\quad 
      + r \bigg(
      D_{(a}h_{b)}' + (F-|h|^2)\gamma_{ab}' + \frac{1}{2}\gamma^{cd}\gamma_{cd}' D_{(a}h_{b)} + \frac{1}{2} D_c (h^c \gamma_{ab}') 
      - h^c (D_{(a}\gamma_{b)c}') \notag\\
      &\qquad\qquad 
      -h_{(a}'h_{b)}
      + h_{(a}\gamma_{b)d}{h^d}' - \gamma_{c(a}' D^c h_{b)} + D_c ( h^c \gamma_{ab}') - \gamma_{c(a}' D^c h_{b)}
      \bigg)
      \notag\\
      &\quad\quad
      + r^2 \bigg[ \frac{1}{4} \gamma^{cd}\gamma_{cd}' (F-|h|^2)\gamma_{ab}' + \frac{1}{2}(F-|h|^2)' \gamma_{ab}' + \frac{1}{2} (F-|h|^2)\gamma_{ab}''\notag\\
      &\qquad\qquad 
      - h_{(a}' \gamma_{b)c}'h^c + \frac{1}{2} h^c h^e \gamma_{ac}' \gamma_{be}' - \frac{1}{2}(F-|h|^2)\gamma_{ac}'\gamma_{bd}' \gamma^{cd} + \frac{1}{2} h_a' h_b'
      \bigg]\ ,
      \label{Rab}
\end{align}
where $(\cdot)'$ represents the derivative with respect to $\partial_r$, $|\cdot| = \sqrt{\langle \cdot , \cdot \rangle_{\gamma}}$ is the magnitude taken by the induced metric $\gamma$, and $D$ is the covariant derivative associated to the metric $\gamma$.
To write out the equations for $\{ R_{v \mu} \} = \{ R_{vv},R_{vr},R_{va} \}$, we denote $\Omega = \dd \T^\flat$, and then we have
\begin{align*}
    \Omega^{ra} 
    &= r^2 [ Fh^a - D^a F - h^c (D_c h^a - D^a h_c)] + r^3 [ h^a (F' - h^c h_c') - (F-|h|^2)\gamma^{ab}h_b']\ ,\\
    \Omega^{ab} 
    &= r (D^a h^b - D^b h^a) + r^2 (h^b \gamma^{ac} h_c' - h^a \gamma^{bc}h_c')\ ,\\
    \nabla_\mu \Omega^{\mu \nu} &= -2 R^{\nu}_{\ \ v}\ .
\end{align*}

Thus the equations of $R_{v \mu},\ \mu=v,r,a$ can be expressed as follows.
\begin{align}
    0 = R_{va}
    & = 
    r h_a R_{rv} - \frac{1}{2} \gamma_{ab} \left[ \frac{1}{\sqrt{\det \gamma}} \partial_r \left( \sqrt{\det \gamma} \Omega^{rb} \right) + D_c \Omega^{cb} \right]\ , \label{Rva} \\
    0 = R_{vv} 
    &= 
    \frac{1}{2} D_a \Omega^{ra} + r^2 ( F - |h|^2 ) R_{rv} + r h^a R_{va}\ ,
    \label{Rvv} \\
    0 = R_{vr}
    &= 
    \frac{1}{2 \sqrt{\det \gamma}} \partial_r \left[ \sqrt{\det \gamma} \left( (r^2 F)' - r h^a (rh_a)' \right) \right] + \frac{1}{2} D^a (r h_a)' \notag\\
    &= 
    \left( F - \frac{1}{2}|h|^2 + \frac{1}{2} D^a h_a \right) \notag\\
    &\qquad 
    + r \left[ \frac{1}{4}\gamma^{cd} \gamma_{cd}' ( 2F - |h|^2 ) + \frac{1}{2}D^a h_a' + F' - h^a h_a' - {h^a}' h_a \right] \notag\\
    &\qquad
    + r^2 \left[
    \frac{1}{4}\gamma^{cd} \gamma_{cd}' (F' - h^a h_a') + \frac{1}{2} F'' - \frac{1}{2} {h^a}' h_a' - \frac{1}{2} h^a h_a''
    \right]\ . \label{Rvr}
\end{align}

Restricting to the connected component $\Hh_i = \{ r=0 \}$ of the horizon, the vacuum Einstein equations are reduced to the following equations:
\begin{align}
    0 = R_{rr} \big|_{r=0}
    &= -\frac{1}{2} \gamma^{ab}\gamma_{ab}'' \big|_{r=0} + \frac{1}{4} \gamma^{cd}\gamma^{bd} \gamma_{ab}' \big|_{r=0} \gamma_{cd}' \big|_{r=0} \ , \label{Rrr-restrict} 
    \\
    0 = R_{ra} \big|_{r=0}
    &= \left( \frac{1}{4} \gamma^{cd}\gamma_{cd}' h_a + \frac{1}{2} D^b \gamma_{ab}' - \frac{1}{2 } D_a (\gamma^{bc}\gamma_{bc}') + h_a' - \frac{1}{2} h^b \gamma_{ab}' \right) \bigg|_{r=0}\ , \label{Rra-restrict} 
    \\
    0 = R_{ab} \big|_{r=0}
    &= \bigg( Ric(\gamma)_{ab} + D_{(a}h_{b)} - \frac{1}{2}h_a h_b \bigg) \bigg|_{r=0}\ , \label{Rab-restrict}    
    \\
    0 = R_{vr} \big|_{r=0}
    &= 
    \left( F - \frac{1}{2}|h|^2 + \frac{1}{2} D^a h_a \right) \bigg|_{r=0}\ ,  \label{Rvr-restrict}
\end{align}
where we have used the degeneracy $\kappa = 0$ in \eqref{Rab-restrict}. The rest equations for the Ricci tensor are trivial on the horizon component.
    \section[Axisymmetric Killing field around 
non-rotating and degenerate
horizon components]
{ Axisymmetric Killing field around \texorpdfstring{\\}{}  non-rotating and degenerate
horizon components}

\label{sec: Axisymmetric Killing field}

This section is divided into two parts.
In Subsection \ref{subsec:3.1}, we construct the axisymmetric Killing field on a non-rotating and degenerate connected component of the horizon. Since a non-rotating and degenerate horizon in a stationary spacetime is a degenerate Killing horizon, a general result that every degenerate Killing horizon must admit an axisymmetric Killing field was first proved by Dunajski--Lucietti in \cite{Lucietti_2023}. For convenience and completeness, we write out the proof in the that subsection. We stress that only a smooth metric is required for the existence of this axisymmetric Killing field on the horizon component.

As subsequent work, in Subsection \ref{subsec:3.2}, we extend this axisymmetric Killing field to the domain of outer communication by the analyticity of the metric. This result also gives affirmatively answers to the questions raised by Lucietti in \cite{Lucietti_2026} that
vacuum transversal derivatives of the extremal Kerr horizon to all orders are in fact unique, and combining the argument in Section \ref{sec:The proof of the main theorem}, under the asymptotically flat condition for the analytic spacetime, the extremal Kerr black hole solution is the unique vacuum spacetime that contains an extremal Kerr Killing horizon. 

\subsection{The axisymmetric Killing field on the horizon}

\label{subsec:3.1}

To find the axisymmetric Killing field on the horizon, in our view, the idea is inspired by an equivalent characterization of the degeneracy of (any connected component of ) the horizon when the horizon is Killing horizon and the inner expansion is non-positive and negative at some point on the cross-section (see, e.g., \cite{Booth-Fairhust}, \cite{Andersson-Mars-Simon}); that is, under appropriate rescaling of the null vectors on the cross-section $S = \{ v=0,r=0 \}$ and the condition stated above, the first order variation of the outer expansion of $S$, with respect to the opposite direction of the inner null vector, is zero (while for non-degenerate horizon, is positive). 

More precisely, let $n = -\partial_r\ ,\ l=\partial_v$ be the two future directed, null normal vectors on $\Hh_i$. Suppose the variation of $S$ along the inner null direction $n$ is $S_s = \{ v=0\ ,\ r=s \psi(x) \}\ ,\ s\in (-\varepsilon,\varepsilon) $, where $\psi=\psi(x)\ ,\ x=(x^1,x^2)$ is a smooth function. Then the linear independent tangent field on $S_s$ becomes $e_a^s = \partial_a + s \psi_a \partial_r$, where $\psi_a = D_a \psi$. The induced metric on $S_s$ is $q_{ab}^s = \gamma_{ab}(s\psi,x) \dd x^a \dd x^b $. Choose the inner null vector $n^s = n$ on $S_s$. Then the outer null vector $l^s$ and its dual one-form $(l^s)^\flat$ become 
\begin{equation}
    \begin{aligned}
        l^s 
        &=
        \partial_v - \left[ \frac{1}{2} (s\psi)^2 (F-|h|^2) + \frac{1}{2} s^2 |\dd \psi|^2 
        \right] \partial_r - ( s\psi h^a + s\psi^a) \partial_a\ ,\\
        (l^s)^{\flat} 
        &=
        \dd r - s \psi_b \dd x^b + O(s^2)\dd v \ ,
    \end{aligned}
\end{equation}
which is determined by $\langle l^s,n \rangle =0\ ,\ \langle l^s,l^s \rangle =0\ ,\ l^s \perp S_s$.
Let $\theta^{(s)} = \theta(l^s)$ be the outer expansion, and $\theta(n)$ be the inner expansion of $S$.
The outer expansion $\theta^{(s)}$ of $S_s$ now takes the form:
\begin{align}
    \theta_{ab}^{(s)} &\overset{\triangle}{=}
    \langle \nabla_{e_a^s} l^s , e_b^s \rangle \notag\\
    &= -s [ D_a D_b \psi +  \psi D_{(a}h_{b)} + h_{(a}D_{b)}\psi ] + O(s^2)\ .\\
    \theta^{(s)} 
    &\overset{\triangle}{=}
    (q^s)^{ab} \theta_{ab}^{(s)} \notag\\
    &=
    -s [ \Delta \psi + h^a D_a \psi + (D_a h^a)\psi ] + O(s^2)\ ,
\end{align}
where $\Delta = \Delta_{\gamma}$ is the Beltrami-Laplacian operator associated with the induced metric $\gamma=\gamma_{ab}(0,x)$.

Therefore the first order of the variation along the null vector $-\psi n$ of the outer expansion $\theta^{(s)}$ is
\begin{align}
    \delta_{-\psi n}\theta^{(s)}
    =
    \frac{d}{ds}\bigg|_{s=0} \theta^{(s)}
    &=
    - [ \Delta \psi + h^a D_a \psi + (D_a h^a)\psi ] \notag\\
    &=
    - D_a (D^a \psi + \psi h^a)\ . \label{deltaxita}
\end{align}
We denote the elliptic operator given from the RHS of \eqref{deltaxita} as 
\begin{align}
    L \psi 
    \overset{\triangle}{=}
    - D_a (D^a \psi + \psi h^a)\ ,
\end{align}
and then by standard functional theory (or, according to Lemma 4.1 of \cite{Andersson-Mars-Simon}), the operator $L$ has positive eigenfunction $\psi > 0$, such that $L\psi = 0$, and $\ker L = span_{\mathbb{R}} \{ \psi \}$. 

Indeed, for arbitrary surface gravity $\kappa \ge 0$ in vacuum spacetime, the stability operator defined in \cite{Andersson-Mars-Simon} is
\begin{equation*}
    L_{stab} \psi 
    \overset{\triangle}{=}
    \delta_{-\psi n}\theta^{(s)}
    = 
    L \psi + \frac{\kappa}{2} ( \tr_{\gamma} \gamma' ) \psi\ ,
\end{equation*}
where $ \frac{1}{2}( \tr_{\gamma} \gamma' ) = - \theta(n)$.
Thus a connected component of the horizon is degenerate, if and only if $L_{stab} \psi = 0$ for some $\psi > 0$ on the cross-section $S$ of that component when the inner expansion $\theta(n)$ satisfies $\theta(n) \le 0 $ and $\theta(n) \not\equiv 0$ on $S$.

In our case, the component $\Hh_i$ is degenerate, so we can avoid to consider the inner expansion $\theta(n)$, and use the operator $L$ directly rather than $L_{stab}$. If we define $\J^a = D^a \psi + \psi h^a$, where $\psi > 0$ is the eigenfunction of the operator $L$, then as stated before, on $S$ we have
\begin{equation}
    D_a \J^a = D_a ( D^a \psi + \psi h^a ) = -L \psi = 0\ . 
    \label{divJ}
\end{equation}

It was first proved by Donajski--Lucietti in \cite{Lucietti_2023} that $\J$ is the axisymmetric Killing field on $S$ that we want. For completeness, we briefly state the proof in the following.
Since $\dim S =2$, let $K$ be the Gauss curvature of $S$, and we can rewrite \eqref{Rab-restrict} as 
\begin{align}
    Ric(\gamma)_{ab} = K \gamma_{ab} 
    &= \frac{1}{2}h_a h_b - D_{(a}h_{b)} \notag \\
    &=
    \frac{1}{2 \psi^2} ( \J_a \J_b - \psi_a \psi_b) - \frac{1}{\psi} B_{ab} + \frac{1}{\psi} D_a D_b \psi\ , 
    \label{Ric(gamma)}
    \\
    K = \frac{1}{2} R(\gamma)
    &= 
    \frac{1}{4 \psi^2} ( |\J|^2 - |d \psi|^2 ) + \frac{1}{2 \psi} \Delta \psi\ ,
    \label{K(gamma)}
\end{align}
where $B_{ab} = D_{(a} \J_{b)}$. Thus by the second Bianchi identity $D^a Ric(\gamma)_{ab} - \frac{1}{2} D_b R(\gamma) = 0$ on $S$, and the formulae deduced from \eqref{Ric(gamma)}, \eqref{K(gamma)}, 
\begin{align}
    D^a Ric(\gamma)_{ab} 
    &= 
    -\frac{1}{2 \psi^3} \left[ (\J^a D_a \psi) \J_b - |\dd \psi|^2 \psi_b \right] 
    + \frac{1}{2 \psi^2} \left[ \J_a D^a \J_b - \Delta \psi \psi_b - \psi_a D^a \psi_b \right] \notag\\
    &\quad\ 
    - \frac{1}{\psi} D^a B_{ab} + \frac{1}{\psi} D_b \Delta \psi\ , 
    \\
    \frac{1}{2} D_b R(\gamma) = D_b K
    &= 
    -\frac{1}{2 \psi^3} \left[ |\J|^2 - |\dd \psi|^2 \right] \psi_b
    + \frac{1}{2 \psi^2} \left[ \J^a D_b \J_a - \psi^a D_b D_a \psi \right] \notag\\
    &\quad\ 
    - \frac{1}{2 \psi^2} \Delta \psi \psi_b + \frac{1}{2 \psi} D_b \Delta \psi\ ,
\end{align}
Subtracting the above two equations, we arrive at the following. 
\begin{align}
    0 
    &= -\frac{1}{2 \psi^3} \left[ (\J^a D_a \psi) \J_b - |\J|^2 \psi_b \right]  + \frac{1}{2 \psi^2} ( \J_a D^a \J_b - \J^a D_b \J_a) \notag \\
    &\quad\ 
    - \frac{1}{\psi} D^a B_{ab} + \frac{1}{2 \psi} D_b \Delta \psi\ . \label{Bianchi-on-S}
\end{align}
Contracting the above equation with $\J^b$, we obtain 
\begin{align}
    \J^b D^a B_{ab} 
    &= D^a (\J^b B_{ab}) - |B|^2 \notag \\
    &=    
    \frac{1}{2} \J^a D_a (\Delta \psi) 
    = \frac{1}{2} D^a (\Delta \psi \J_a)
    \ , 
    \label{B1}
    \\
    |B|^2 
    &= D^a \left( \J^b B_{ab} - \frac{1}{2} \Delta \psi \J_a \right)\ . \label{B2}
\end{align}
The last equality of \eqref{B1} have used \eqref{divJ}. Thus integrate \eqref{B2} over $S$ and by the divergence theorem, we have $|B|^2 \equiv 0$ on $S$, and thus 
\begin{equation}
    ( \lie_{\J} \gamma )_{ab} = B_{ab} = 0\ ,
    \label{LieJ-gamma=0}
\end{equation}
that is, $\J$ is a Killing field on $S$.

$\J$ cannot vanish everywhere on $S$. Otherwise, $0 = \J = D \psi + \psi h$, and $h = - \dd \ln \psi$, which is the key ingredient of the non-existence of non-rotating and degenerate horizon components in static vacuum spacetimes in \cite{Chrusciel-degenerate}, as it will lead to a contradiction with the Gauss-Bonnet theorem.
In more detail, if $h = - \dd \ln \psi$, then the Ricci equation \eqref{Rab-restrict} can be simplified as the equation
\begin{equation}
    \label{hclosed-u}
    \frac{1}{2} K u \gamma_{ab}
    =
    D_a D_b u\ ,
\end{equation}
where $u = \sqrt{\psi}$.
Taking the trace of \eqref{hclosed-u} and differentiating it once with the index $a$ contracted, we have
\begin{align}
    Ku 
    &= \Delta u\ ,
    \label{Ku} \\
    \frac{1}{2} D_b (Ku) 
    &= D^a D_a D_b u
    \notag \\
    &= D_b \Delta u + R^{a}_{\ \ bac}(\gamma) D^c u
    \notag \\
    &= D_b \Delta u + Ric(\gamma)_{bc} D^c u
    \notag \\
    &= D_b \Delta u + K D_b u\ .
    \label{D(Ku)}
\end{align}
Substitute \eqref{Ku} into \eqref{D(Ku)}, and we have 
\begin{equation}
    u \dd K + 3 K \dd u = 0\ ,
\end{equation}
which is equivalent to $Ku^3 \equiv C$ on $S$, where $C$ is a constant. Thus by the compactness of $S$,
\begin{equation}
    0 = 
    \int_S \Delta u = \int_S Ku = \int_S \frac{C}{u^2}\ ,
\end{equation}
which forces $C = 0$ and thus $K = 0$ on $S$. However, the Gauss-Bonnet theorem tells that $\int_S K = 4 \pi$, which yields a contradiction to the above result.
Therefore, $\J \not\equiv 0$ on $S$. 

Since $\J$ is a Killing field on $S$, $D_a \J_b$ is antisymmetric. Hence $( D_a \J_b )$ must be either zero or of rank two at any point of the two-dimensional cross-section $S$. In particular, consider any zero point $p \in S$ of $\J$. If $ D_a \J_b (p) = 0$, then for every geodesic $\eta$ emanating from $p$, by the Killing equation $D_{\dot\eta} D_{\dot\eta} \J = -R(\dot\eta, \J ) \dot\eta $ with the initial data $J(p) = 0 , D_{\dot\eta} \J (p) = 0$ , we have $\J \equiv 0$ along $\eta$. By the arbitrariness of the choice of the geodesic $\eta$, we conclude that $\J \equiv 0$ everywhere on $S$, which leads to a contradiction. Therefore, $\det (D_a \J_b) (p) \neq 0$, which means that the zero point $p$ is non-degenerate.
By the Hopf-Poincar\'e lemma, $\J$ must have and only have two zero points on $S$. Again, since $\J$ is a Killing field on $S$, the distance induced from $\gamma$ on $S$ between any point of $S$ and (one of) the zero point of $\J$ is preserved under the flow generated by the vector field $\J$. Thus the integral curves of $\J$ must be spacelike closed curves on $S$.

Now by \eqref{B1} and $B_{ab} = 0$, we have $0 = \J^a D_a (\Delta \psi) = \Delta ( D_{\J} \psi)$. Applying the maximum principle and exploiting the compactness of $S$, $D_{\J} \psi$ is a constant on $S$.
Note that $\J$ has zero points on $S$, and we obtain
\begin{align}
    D_{\J} \psi = 0\ .
    \label{LieJ-psi=0}
\end{align}

By the definition of $\J$, $h_a = \frac{1}{\psi} ( \J_a - D_a \psi)$. Since $\J$ is a Killing field, $[ \lie_{\J}, D] =0$ on $S$. Therefore due to \eqref{LieJ-psi=0}, we have
\begin{align}
    (\lie_{\J} h)_a = 0\ .
    \label{LieJ-h=0}
\end{align}

Finally, by the restricted equation \eqref{Rvr-restrict}, we have
\begin{align}
    \lie_{\J} F = \lie_{\J} \left(  \frac{1}{2}|h|^2 - \frac{1}{2} D^a h_a \right) = 0\ .
    \label{LieJ-F=0}
\end{align}

Note that the above results \eqref{LieJ-gamma=0},\eqref{LieJ-psi=0} \eqref{LieJ-h=0},\eqref{LieJ-F=0} are independent of the coordinate $v$, we can directly lift the vector field $\J$ to the whole $\Hh_i$ by Lie dragging with respect to the stationary Killing field $\T = \partial_v$, which is still denoted as $\J = \J^a(x) \partial_a$ on $\Hh_i$, such that \eqref{LieJ-gamma=0}, \eqref{LieJ-psi=0},\eqref{LieJ-h=0},\eqref{LieJ-F=0} still hold. As in \cite{Hawking-Ellis}, one can first choose a shortest geodesic joining the two zero points of $\J$ on $S$, and denote the tangent vector as $Y_4$, and then extend $Y_4$ to be a vector field on $S$ by Lie dragging with respect to $\J$, that is, $\lie_{\J} Y_4 = 0$ with $Y_4 = 0$ on the two zero points of $\J$ and coinciding the value on that geodesic. $Y_4$ also can be lifted to $\Hh_i$ by Lie dragging with respect to $\T$. Suppose the smallest positive period of $\J$ is $t_1$, and then we can construct a new coordinate system, such that $\{ v, x^1,x^2 \} = \{ v, \phi, \theta \}$, and $Y_1 \overset{\triangle}{=} \T = \partial_v, Y_3 \overset{\triangle}{=} \J = \frac{2 \pi}{t_1} \partial_{\phi}, Y_4 = \partial_{\theta}$ as in \cite{Hawking-Ellis}, which form a coordinate basis on $\Hh_i$.

As a consequence, we obtain an axisymmetric Killing field $\J$ defined on the entire horizon component $\Hh_i = \{ r=0 \}$, that is,
\begin{align}
    (\lie_{\J} g)_{ab} \big|_{r=0} = 0\ .
    \label{LieJ-g=0}
\end{align}

\subsection{Extension of the axisymmetric Killing field}

\label{subsec:3.2}

At this point, we have found an axisymmetric Killing field $\J$ on $\Hh_i$ satisfying \eqref{LieJ-g=0}, which is equivalent to the following conditions:
\begin{align}
    (\lie_{\J} \gamma)\big|_{r=0} = 0,\ 
    (\lie_{\J} h)\big|_{r=0} = 0,\
    (\lie_{\J} F)\big|_{r=0} = 0,\ 
    \label{initial-lie-data}
\end{align}

Using a method similar to that in \cite{Hawking-Ellis}, we can extend $\J = \frac{2 \pi}{t_1} \partial_{\phi}, Y_4 = \partial_{\theta}$ to be vector fields in a neighbourhood $\mathcal{U}$ of $\Hh_i$ by Lie dragging with respect to $Y_2 \overset{\triangle}{=} \partial_r$. Since $\lie_{\J} T = 0$ on $\Hh_i$ and $[\J,\partial_r] = 0$ in the open set $\mathcal{U} \subset \M$, we have $\lie_{\partial_r } \lie_{\J} T = \lie_{\J} \lie_{\partial_r } T = 0$ in $\mathcal{U}$, and hence $\lie_{\J} T \equiv 0$ in $\mathcal{U}$. The same holds for $Y_4$, and therefore we obtain a coordinate system $\{ v,r,x^1,x^2 \} = \{ v,r,\phi,\theta \}$ on $\mathcal{U}$, such that $\{ Y_1 = \T = \partial_v, Y_2 = \partial_r, Y_3 = \J = \frac{2 \pi}{t_1} \partial_{\phi}, Y_4 = \partial_{\theta} \}$ becomes a coordinate basis in $\mathcal{U}$.

Therefore, by the analyticity of the metric $g$, if the Lie derivative of any transversal derivatives of $g$ with respect to $\J$ vanishes, then $\J$ can be extended to the whole region of the domain of the outer of communications $\D$. To show this, we will argue by induction.

Denote $(g_k)_{ab} = \partial_r^k \big|_{r=0} g_{ab}\ ,\ (\gamma_k)_{ab} = \partial_r^k \big|_{r=0} \gamma_{ab} \ ,\ (h_k)_{a} = \partial_r^k \big|_{r=0} h_{a}\ , F_k = \partial_r^k \big|_{r=0} F$, for $ k=0,1,2,\dots$. What we want to show is $\lie_{\J} \gamma_k = 0\ ,\ \lie_{\J} h_k = 0\ ,\ \lie_{\J} F_k = 0 $, for $k = 1,2,\dots$.

However, when considering the first order of the transversal derivative of $\gamma$, one will find such a necessary condition on $S$: if $\lie_{\J} \gamma_1 = 0$ on $S$, then we have $\tr_{\gamma} \lie_{\J} \gamma_1 = \lie_{\J} (\tr_{\gamma} \gamma_1 ) = 0$ on $S$, which does not hold \textit{a priori}. Thus we have to choose a new cross-section $S^f$ to realize this condition. 

Suppose that $S^f \overset{\triangle}{=} \{ v=f(x)\ ,\  r=0 \}$, where $f=f(x)$ is a smooth function. The linear independent basis on $S^f$ now becomes $ e_a^f = \partial_a + f_a T $, where $f_a = D_a f$, and $a=1,2$. Repeat the same argument for the surface $S$, there is still an axisymmetric Killing field $\J^f$ on $S^f$, and $ (\J^f)_a = D_a \psi + \psi h_a$, where $\psi > 0$ is a positive eigenfunction of the operator $L$ on the smooth compact two-surface $S^f$. Thus $\J^f = \J^a e_a^f = \J + (D_{\J}f ) T$.
We still have $\gamma^f_{ab} = \langle e_a^f , e_b^f \rangle_{g} = \gamma_{ab}$, and
$ ( \lie_{\J^f} \gamma^f )_{ab} = 0 $.

One can calculate that
\begin{align}
    (\gamma_1^f)_{ab} 
    &=
    \partial_r \gamma_{ab}^f 
    = - (\lie_{n} g)_{ab} 
    = -2 \langle \nabla_{e_a^f} n , {e_b^f} \rangle 
    \notag\\
    &= (\gamma_1)_{ab} - 2 [ D_a D_b f - h_{(a}D_{b)}f ]\ ,
\end{align}
and therefore $\tr_{\gamma^f} \gamma_1^f = \tr_{\gamma} \gamma_1^f = \tr_{\gamma} \gamma_1 - 2 [ \Delta_{\gamma} f - h^a D_a f ]$. To satisfy the necessary condition $\lie_{\J} ( \tr_{\gamma^f} \gamma_1^f) = 0$ on $S^f$, we can set $ \tr_{\gamma^f} \gamma_1^f =\tr_{\gamma} \gamma_1 - 2 [ \Delta_{\gamma} f - h^a D_a f ] \equiv c$ on $S^f$, where $c$ is a constant. Thus we have the following equation for the function $f$:
\begin{align}
    Pf
    \overset{\triangle}{=}
    \Delta_{\gamma} f - h^a D_a f = \frac{1}{2} ( \tr_{\gamma} \gamma_1 - c )\ .
    \label{choose-f}
\end{align}
Since the operator $P=-L^*$,
by standard functional theory, this equation has a solution in $H^1 (S)$ if and only if $ \tr_{\gamma} \gamma_1 - c \in (\ker P^* )^{\perp} = ( \ker L)^{\perp} = ( span_{\mathbb{R}
} \{ \psi \} )^{\perp} $. Thus if we choose 
\begin{align}
    c = \frac{ \int_{S} \psi (\tr_{\gamma} \gamma_1) \dd A}{\int_{S} \psi \dd A }\ ,
    \label{choose-c}
\end{align}
then the equation \eqref{choose-f} has a solution $f \in H^1(S)$. Since $\gamma\ ,\ h\ ,\ \gamma_1$ are at least smooth, by standard PDE regularity theory, this solution $f$ is smooth. Therefore, without loss of generality, we may assume from the outset that $S = S^f$, and there is a Killing field $\J$ on $\Hh_i$, such that
\begin{align}
    (\lie_{\J} \gamma ) \big|_{r=0} = 0\ ,\ (\lie_{\J} h ) \big|_{r=0} = 0\ ,\ (\lie_{\J} F ) \big|_{r=0} = 0\ ,
\end{align}
and at the same time,
\begin{align}
    \tr_{\gamma} \gamma_1 \equiv c\ ,\ 
    \lie_{\J} ( \tr_{\gamma} \gamma_1 ) = 0 \quad \text{on}\  S\ .
\end{align}

Now we start the induction.
For any $k \ge 1$, suppose that $\lie_{\J} \gamma_j = 0\ ,\   \lie_{\J} h_j = 0\ ,\  \lie_{\J} F_j = 0$, for any $0 \le j < k$.

Then for the $k$-th order, denote $Q_k = \lie_{\J} \gamma_k$, and the Taylor series now reads
\begin{align}
    \lie_{\J} \gamma(r) 
    &= \frac{r^k}{k!} \lie_{\J} \gamma_k + O(r^{k+1})
    = \frac{r^k}{k!}Q_k + O(r^{k+1})\ ,
    \label{lie-gammak}
    \\
    \lie_{\J} h(r)
    &= \frac{r^k}{k!} \lie_{\J} h_k + O(r^{k+1})\ ,
    \label{lie-hk}
    \\
    \lie_{\J} F (r)
    &= \frac{r^k}{k!} \lie_{\J} F_k + O(r^{k+1})\ .
    \label{lie-Fk}
\end{align}
One may also calculate that
\begin{align}
    (\lie_{\J} \gamma^{ab} (r) ) = - \gamma^{ac} \gamma^{bd} \lie_{\J} \gamma_{cd}
    =
    - \frac{r^k}{k!} (Q_k)^{ab} + O(r^{k+1})\ .
\end{align}

For convenience, we will omit the subscript $k$ of $Q_k$ in what follows. 

\textbf{Step 1}: When $k = 1$, by the choice of the cross-section $S$ as stated before, $\tr_{\gamma} Q = \tr_{\gamma} \lie_{\J} \gamma_1 = \lie_{\J} ( \tr_{\gamma} \gamma_1 ) = 0 $ on $S$.

When $k > 1$, since $R_{rr} \equiv 0$, we have $\lie_{\J} R_{rr} \equiv 0$. Substitute \eqref{lie-gammak} and \eqref{Rrr} into this equation, we obtain
\begin{align}
    0 = \lie_{\J} R_{rr}
    = -\frac{r^{k-2}}{2 (k-2)!} \gamma^{ab} Q_{ab} + O(r^{k-1})\ ,
\end{align}
and thus $\tr_{\gamma} Q = \gamma^{ab} Q_{ab} = 0$ on $S$.

\textbf{Step 2}: Denote the covariant derivative associated to the metric $\gamma_{ab}(r) = \gamma_{ab}(r,x^1,x^2)$ by $D^{(r)}_a$. Then we have
\begin{align}
    \lie_{\J} ( D_{(a}^{(r)} h_{b)} (r) ) 
    = 
    \frac{r^k}{k!}
    \left[ D_{(a}( \lie_{\J}h_k)_{b)} - h^c D_{(a}Q_{b)c} + \frac{1}{2} h^c D_c Q_{ab} \right] + O(r^{k+1})\ . 
\end{align}
Meanwhile, we can calculate that
\begin{align}
    \lie_{\J} h^a (r)
    = 
    \lie_{\J} ( h_b (r) \gamma^{ab}(r) )
    &=
    \frac{r^k}{k!}\gamma^{ab} ( \lie_{\J} h_k)_b - \frac{r^k}{k!} h_b Q^{ab} + O(r^{k+1}) \notag\\
    &=
    \frac{r^k}{k!} \left[ ( \lie_{\J} h_k)^a - h_b Q^{ab} \right] + O(r^{k+1})\ .
\end{align}

Since $ R_{ra} \equiv 0$, we have $\lie_{\J} R_{ra} \equiv 0$, and thus by \eqref{Rra}, we have
\begin{align}
    0 = \lie_{\J} R_{ra}
    =
    \frac{r^{k-1}}{2 (k-1)!} \left[ (k+1) (\lie_{\J} h_k)_a - k h^b Q_{ab} + D^b Q_{ab}
    \right] + O(r^k)\ ,
\end{align}
where we have used \textbf{Step 1} ($\tr_{\gamma}Q = 0$) to simplify the result.
Thus $\lie_{\J}h_k$ is determined by $Q$,
\begin{align}
    ( \lie_{\J}h_k )_a
    =
    \frac{k h^b Q_{ab} - D^b Q_{ab}}{k+1}\ ,
    \label{lie-hk2}
\end{align}
and if $ Q = 0$, then we have $\lie_{\J}h_k = 0$ on $S$.

\textbf{Step 3}: Since $R_{rv} \equiv 0$, we have $ \lie_{\J} R_{rv} \equiv 0 $, and substitute \eqref{lie-gammak} to \eqref{lie-Fk} and \eqref{Rvr} into this equation, we obtain
\begin{align}
    0 
    &=
    \lie_{\J} R_{rv} \notag\\
    &=
    \frac{r^{k}}{k!} 
    \bigg[ (k+1)(k+2) \lie_{\J}F_k - (k+1)(k+2) h^a ( \lie_{\J} h_k )_a 
    + (k+1) h^a h^b Q_{ab} 
    \notag\\
    &\qquad\quad
    + (k+1)D^a (\lie_{\J} h_k)_a - (D^a h^b) Q_{ab} - h^a D^b Q_{ab} \bigg] + O(r^{k+1})\ ,
\end{align}
where we have used \textbf{Step 1} ($\tr_{\gamma}Q = 0$) to simplify the result. From the above equation, we have
\begin{align}
    \lie_{\J}F_k 
    =
    h^a ( \lie_{\J} h_k )_a 
    - \frac{h^a h^b Q_{ab} + D^a (\lie_{\J} h_k)_a}{k+2}
    + \frac{(D^a h^b) Q_{ab} + h^a D^b Q_{ab}}{(k+1)(k+2)}\ .
    \label{lie-Fk2}
\end{align}
Again, if $ Q = 0 $, then by \textbf{Step 2} and the above equation, we have $\lie_{\J} F_k = 0 $ on $S$.

As a consequence, to complete the induction, we only need to prove that $Q = 0$.

\textbf{Step 4}: Since $R_{ab} \equiv 0 $, we have $\lie_{\J} R_{ab} \equiv 0$. Substitute \eqref{lie-gammak} to \eqref{lie-Fk} and \eqref{Rab} into this equation, we obtain
\begin{align}
    0 
    &= 
    \lie_{\J} R_{ab} \notag\\
    &=
    \frac{r^k}{k!} \Bigg[
    \frac{1}{2} D^c D_a Q_{bc} + \frac{1}{2} D^c D_b Q_{ac} - \frac{1}{2} \Delta Q_{ab} 
    \notag\\
    &\qquad\qquad
    + (k+1) \left( D_{(a} (\lie_{\J} h_k )_{b)} - h^c D_{(a}Q_{b)c} + \frac{1}{2} h^c D_c Q_{ab} \right)
    \notag\\
    &\qquad\qquad 
    - (k+1) ( \lie_{\J} h_k )_{(a}h_{b)}
    + \frac{k(k+1)}{2}(F - |h|^2) Q_{ab}
    \notag\\
    &\qquad\qquad 
    + k h_{(a}Q_{b)c} h^c + \frac{k}{2} ( \ddiv_{\gamma} h ) Q_{ab} + \frac{k}{2} h^c D_c Q_{ab}
    - k Q_{c(a}D^c h_{b)}
    \bigg]
    + O(r^{k+1})\ .
    \label{lie-Rab1}
\end{align}
Thus substitute \eqref{lie-hk2} into \eqref{lie-Rab1}, and note that 
\begin{align*}
    (k+1) D_{(a}(\lie_{\J}h_k)_{b)}
    &=
    k (D_{(a}h^c )Q_{b)c} + k h^c D_{(a}Q_{b)c} - D_{(a}D^c Q_{b)c}\ ,
    \notag\\
    -(k+1) (\lie_{\J}h_k)_{(a}h_{b)}
    &=
    - k h_{(a}h^c Q_{b)c} + h_{(a}D^c Q_{b)c}\ ,
\end{align*}
we have
\begin{align}
    0
    &=
    \frac{1}{2} \bigg(
    D^c D_a Q_{bc} + D^c D_b Q_{ac} - \Delta Q_{ab} - D_a D^c Q_{bc} - D_b D^c Q_{ac}
    \bigg)
    \notag\\
    &\qquad\ 
    + \frac{2k+1}{2} h^c D_c Q_{ab} - h^c D_{(a}Q_{b)c} + h_{(a}D^c Q_{b)c}
    + k (D_{(a} h^c - D^c h_{(a} )Q_{b)c}
    \notag\\
    &\qquad\ 
    + \left[
    \frac{k(k+1)}{2}(F-|h|^2) + \frac{k}{2} \ddiv_{\gamma} h
    \right] Q_{ab}\ .
    \label{lie-Rab2}
\end{align}
Also note that by the definition of Riemannian curvature tensor,
\begin{align*}
    D^c D_a Q_{bc} - D_a D^c Q_{bc}
    &=
    R^{c \ \ \ d}_{\ ab} Q_{dc} + R^{c \ \ \ d}_{\ ac} Q_{bd}
    \\
    &=
    K(\gamma^{c}_{\ \ b} \gamma_{a}^{\ \ d} - \gamma^{cd} \gamma_{ab})Q_{dc} + R_{a}^{\ \ d}Q_{bd}
    \\
    &= 2K Q_{ab}\ ,
\end{align*}
where we have used \textbf{Step 1} ($\tr_{\gamma}Q = 0$) in the last equality. Thus \eqref{lie-Rab2} can be simplified as
\begin{align}
    0
    &=
    -\frac{1}{2} \Delta Q_{ab} + \frac{2k+1}{2} h^c D_c Q_{ab} + (T_k Q)_{ab}
    + \left[ 2K + \frac{k(k+1)}{2}(F - |h|^2) + \frac{k}{2} \ddiv_{\gamma} h \right] Q_{ab}
    \notag\\
    &=
    -\frac{1}{2} \Delta Q_{ab} + \frac{2k+1}{2} h^c D_c Q_{ab} + (T_k Q)_{ab}
    - \left[
    \frac{(k-1)(k+2)}{2}F + \frac{k^2 +1}{2} \ddiv_{\gamma} h
    \right] Q_{ab}\ ,
    \label{lie-Rab3}
\end{align}
where we have used $2K = F - \frac{1}{2} \ddiv_{\gamma} h\ ,\ |h|^2 = 2F +\ddiv_{\gamma} h $ from \eqref{Rab-restrict} and \eqref{Rvr-restrict}, and
the operator $T_k$ is defined on the space of the traceless symmetric two tensors as
\begin{align}
    T_k Q_{ab} \overset{\triangle}{=}
    - h^c D_{(a}Q_{b)c} + h_{(a}D^c Q_{b)c}
    + k (D_{(a} h^c - D^c h_{(a} )Q_{b)c}\ .
\end{align}

Here we point out that, in the case where the horizon component is \textit{rotating}, one can also calculate the Lie derivative of the transversal derivatives of $\gamma$ with respect to the candidate Killing field by restricting the Einstein equations on the horizon (see, e.g., \cite{Hollands-Ishibashi-Wald2007},\cite{Hollands-Ishibashi}), rather than using the NP formulae \cite{Hawking-Ellis}. 
In the rotating case, the argument is to decompose the stationary Killing field $\T$ as $\K \overset{\triangle}{=} \T -\Ll$ and $\Ll$, where $\Ll$ is a null vector tangent to the generators of the horizon component, and $\K$ is expected to be the axisymmetric Killing field. Thus to show that $\K$ is a Killing field, it suffices to prove that $\Ll$ is a Killing field. One can also introduce the Gaussian null coordinates with $\Ll = \partial_v$, and then the equations for $R_{ab}$ can be expressed as follows.
\begin{align}
    0 = R_{ab} \big|_{r=0}
    &= \bigg( - \partial_v \gamma_{ab}' + Ric(\gamma)_{ab} + D_{(a}h_{b)} - \frac{1}{2}h_a h_b \bigg) \bigg|_{r=0}\ , 
    \label{Rab-restrict-rotating}   
\end{align}
where $(\cdot)' $ still represents the derivative with respect to $\partial_r$. The extra term $- \partial_v \gamma_{ab}'$ in the parentheses, compared to \eqref{Rab-restrict}, substantially simplifies the entire induction argument. For example, after proving that $\Ll$ is Killing on the horizon component, differentiating \eqref{Rab-restrict-rotating} once with respect to $\Ll$, one may directly obtain
\begin{align}
    \lie_{\Ll} Q_1
    \overset{\triangle}{=}
    \lie_{\Ll} (\partial_v \gamma_{ab}') \bigg|_{r=0} = 0 \ ,
\end{align}
where we still use the the notation $Q_k \overset{\triangle}{=} (\lie_{\Ll} \partial_r^k \gamma) \big|_{r=0} 
\overset{\triangle}{=} 
\lie_{\Ll} \gamma_k$ for $k=1,2,\dots$. Therefore, by integrating the inner product of $\lie_{\Ll} Q_1$ and $\gamma_1$ with the metric $\gamma$, one can obtain
\begin{align}
    0 
    &=
    \int_S \langle \gamma_1 , \lie_{\Ll} Q_1 \rangle \dd A
    \notag\\
    &= 
    \int_S \lie_{\Ll} \left( \frac{1}{2} \lie_{\Ll} |\gamma_1|^2 \right) \dd A - \int_S \langle \lie_{\Ll} \gamma_1 \ ,\ Q_1 \rangle \dd A
    \notag\\
    &=
    - \int_S \langle \lie_{\Ll} \gamma_1 \ ,\ Q_1 \rangle \dd A 
    \notag \\
    &= 
    - \int_S |Q_1|^2 \dd A \ ,
\end{align}
and thus $Q_1 = \lie_{\Ll} \gamma_1 = 0$ on $S$. Similarly, the argument for $Q_k = 
\lie_{\Ll} \gamma_k$, the Lie derivative of the $k$-th order transversal derivatives of $\gamma$ with respect to $\Ll$, can proceed by differentiating the equation \eqref{Rab-restrict-rotating} $k$ times with respect to $\Ll$ as above.

However, returning to our case in which the horizon component is \textit{non-rotating}, the extra term in \eqref{Rab-restrict-rotating} vanishes (see \eqref{Rab-restrict}), since the horizon component $\Hh_i$ itself is a Killing horizon. We have to consider the $k$-th variation of the Ricci tensor $R_{ab}$ and consider the elliptic equation \eqref{lie-Rab3} for $Q_{ab}$. Likewise, one may expect that $\lie_{\J} Q = 0$. 
Note that 
\begin{align}
    \frac{1}{2} D^c (\psi^{2k+1}) + \frac{2k+1}{2}\psi^{2k+1} h^c = \frac{2k+1}{2} \psi^{2k+1} \J^c\ .
    \label{weight}
\end{align}
Therefore, contracting
\eqref{lie-Rab3} with $ \psi^{2k + 1} \lie_{\J} Q^{ab}$, and 
then integrating over $S$,  we obtain
\begin{align}
    0
    &=
    \int_{S} \psi^{2k+1} \langle -\frac{1}{2} \Delta Q + \frac{2k+1}{2} h^c D_c Q\  , \ \lie_{\J}Q \rangle \dd A
    + \int_{S} \psi^{2k+1} \langle T_k Q\ ,\ \lie_{\J}Q \rangle \dd A
    \notag\\
    &\qquad\qquad
    - \int_{S} \psi^{2k+1} \left[
    \frac{(k-1)(k+2)}{2}F + \frac{k^2 +1}{2} \ddiv_{\gamma} h
    \right] \langle Q\ ,\ \lie_{\J} Q \rangle \dd A
    \notag\\
    &:=
    I + II + III
    \ .
    \label{lie-Rab4}
\end{align}

Using \eqref{weight}, the integral $I$ can be simplified as follows.
\begin{align}
    I 
    &= 
    \int_{S} \psi^{2k+1} \langle -\frac{1}{2} \Delta Q + \frac{2k+1}{2} h^c D_c Q\  , \ \lie_{\J}Q \rangle \dd A
    \notag\\
    &=
    \frac{2k+1}{2} \int_{S} \psi^{2k} D^a \psi 
    \langle D_a Q\ ,\ \lie_{\J}Q \rangle \dd A
    + \frac{1}{2} \int_{S} \psi^{2k+1} \langle D Q\ ,\ D \lie_{\J}Q \rangle \dd A
    \notag\\
    &\quad
    + \frac{2k+1}{2} \int_{S} \psi^{2k+1}h^a \langle D_{a} Q\ ,\ \lie_{\J}Q \rangle \dd A
    \notag\\
    &=
    \frac{2k+1}{2} \int_{S} \psi^{2k} \langle D_{\J} Q\ ,\ \lie_{\J}Q \rangle \dd A\ ,
    \label{I}
\end{align}
where we have used the fact that the second integral in the second equality vanishes:
\begin{align}
    \int_S \psi^{2k+1} \langle D Q\ ,\ D \lie_{\J}Q \rangle \dd A
    &=
    \frac{1}{2} \int_S \psi^{2k+1} D_{\J} (|DQ|^2)
    \notag\\
    &= 
    -\frac{1}{2} \int_S |DQ|^2 D_a ( \psi^{2k+1} \J^a )
    \notag\\
    &= 0\ ,
\end{align}
where we have used the properties $D_{\J}\psi = 0$ and $\ddiv_{\gamma} \J = D_a \J^a = 0$ on $S$.

The integral $III$ indeed vanishes: for any smooth function $G$ on $S$, if $\lie_{\J}G = 0$, then
\begin{align}
    \int_S G \langle Q\ ,\ \lie_{\J} Q \rangle \dd A
    &=
    \frac{1}{2} \int_S D_{\J} ( G |Q|^2 )
    \notag\\
    &=
    \frac{1}{2} \int_S D_a ( G|Q|^2 \J^a )
    -\frac{1}{2} \int_S G |Q|^2 D_a \J^a 
    = 0\ .
\end{align}
Choose $G = \psi^{2k+1} \left[
    \frac{(k-1)(k+2)}{2}F + \frac{k^2 +1}{2} \ddiv_{\gamma} h
    \right] $, and thus we have $III=0$.

To calculate the integral $II$, note that the dual operator $T_k^*$ of $T_k$ takes the form
\begin{align}
    T_k^* Q_{ab}
    &=
    - h^c D_{(a}Q_{b)c} + h_{(a}D^c Q_{b)c}
    \notag\\
    &\quad\ 
    - (k+1)(D_{(a}h^c - D^c h_{(a})Q_{b)c}
    \notag\\
    &\quad\ 
    + (2k+1)( h_{(a}D^c \ln \psi - h^c D_{(a} \ln \psi)Q_{b)c}\ ,
\end{align}
and thus
\begin{align}
    ( T_k - T_k^* )Q_{ab}
    &=
    (2k+1) \left[
    D_{(a}h^c - D^c h_{(a}
    + h^c D_{(a} \ln \psi - h_{(a}D^c \ln \psi
    \right] Q_{b)c}
    \notag\\
    &=
    \frac{2k+1}{\psi} \left[
    D_{(a}(\psi h^c) - D^c (\psi h)_{(a} 
    \right] Q_{b)c}
    \notag\\
    &=
    \frac{2k+1}{\psi} \left[
    D_{(a}(\J^c) - D^c (\J)_{(a} 
    \right] Q_{b)c}
    \notag\\
    &=
    \frac{2k+1}{\psi} \left[
    (D_a \J^c )Q_{bc}
    + (D_b \J^c)Q_{ac} 
    \right]\ .
\end{align}
Therefore, the integral $II$ can be simplified as
\begin{align}
    II
    &=
    \int_S \psi^{2k+1} \langle T_k Q\ ,\ \lie_{\J}Q \rangle \dd A
    \notag\\
    &=
    \int_S \psi^{2k+1} \langle Q\ ,\ T_k^* \lie_{\J} Q \rangle \dd A
    \notag\\
    &=
    - \int_S \psi^{2k+1} \langle \lie_{\J}Q \ ,\ T_k^* Q \rangle \dd A
    \notag\\
    &= \frac{1}{2} \int_S \psi^{2k+1} \langle (T_k - T_k^*)Q \ ,\ \lie_{\J}Q \rangle
    \notag\\
    &=
    \frac{2k+1}{2} \int_S 
    \psi^{2k} \left[
    (D_a \J^c )Q_{bc}
    + (D_b \J^c)Q_{ac} 
    \right] \lie_{\J}Q^{ab} \dd A\ .
    \label{II}
\end{align}
The introduction of the $T_k^*$ operator to simplify the integral $II$ is suggested by AI.

Substituting \eqref{I},\eqref{II} and $III = 0$ into \eqref{lie-Rab4}, we have
\begin{align}
    0 = I + II + III
    =
    \frac{2k+1}{2} \int_S \psi^{2k} \langle \lie_{\J} Q\ ,\ \lie_{\J} Q \rangle \dd A\ ,
\end{align}
from which we obtain $\lie_{\J}Q = 0$ on $S$. Therefore, integrating the inner product of $\lie_{\J} Q$ and $\gamma_k$ on $S$, we have
\begin{align}
    0 =
    \int_S \langle \gamma_k \ ,\ \lie_{\J} Q \rangle \dd A 
    &=
    \int_S \lie_{\J} \left( \frac{1}{2} \lie_{\J} |\gamma_k|^2 \right) \dd A - \int_S \langle \lie_{\J} \gamma_k \ ,\ Q \rangle \dd A
    \notag\\
    &=
    -\int_S |Q|^2 \dd A\ ,
\end{align}
which means that $Q = \lie_{\J} \gamma_k \equiv 0 $ on $S$, and thus we complete the induction.

As a consequence, $(\lie_{\J}g_k)\big|_{r=0} = 0$. Since the spacetime is $I^+$-regular, the domain of outer communication $\D$ is simply connected (see, e.g., \cite{Galloway1995} or Corollary 2.4 of \cite{Chrusciel2008}). Thus by the analyticity of the metric $g$, $\J$ can be extended to be an axisymmetric Killing field in the domain of outer communication of the spacetime.

    \section{The proof of the main theorem}

\label{sec:The proof of the main theorem}

As stated in the introduction, to prove the \textbf{Main Theorem}, we are going to show that the spacetime we consider is indeed stationary and axisymmetric in the sense of Definition \ref{def:stationary and axisymmetric}, and then by Theorem \ref{thm:uniq-single-bh} we will obtain a contradiction.

According to the construction of the axisymmetric Killing field, $\J$ commutes with $\T$ in the  neighbourhood $\mathcal{U}$ of the Gaussian null coordinates of the non-rotating and degenerate horizon component $\Hh_i$.  Suppose $\K = [\J,\T]$ in the domain of outer communication $\D$. Thus the construction of $\J$ implies that $\K = 0$ in $\mathcal{U}$. Furthermore, we have $\lie_{\K}g = \lie_{\J} \lie_{\T} g - \lie_{\T} \lie_{\J} g = 0$ in $\D$, which means that $\K$ is still a Killing field in $\D$. Note that
for any point $p \in \mathcal{U} \cap \D$, we have $\K (p) = 0\ ,\ \nabla \K (p) = 0$, and $\D$ is simply connected by the $I^+$-regularity, and thus the Killing field $\K$ vanishes everywhere in $\D$, which means that $\J$ indeed commutes with $\T$ everywhere in $\D$.

Therefore, to ensure that the spacetime is stationary and axisymmetric in the sense of Definition \ref{def:stationary and axisymmetric}, it remains to verify the Frobenius conditions
\begin{align}
    \T^{\flat} \wedge \J^{\flat} \wedge \dd \T^{\flat} = 0\ ,\ \T^{\flat} \wedge \J^{\flat} \wedge \dd \J^{\flat} = 0\ 
\end{align}
hold in $\D$.

In fact, since $\J$ is a Killing field in the spacetime, we have the closed form
\begin{align}
    \dd * ( \J^{\flat} \wedge \dd \J^{\flat} ) = 0\ .
\end{align}
Hence by Cartan's formula, we have
\begin{align}
    \dd *( \T^{\flat} \wedge \J^{\flat} \wedge \dd \J^{\flat} ) 
    &=
    \dd \iota_{\T} *( \J^{\flat} \wedge \dd \J^{\flat} )
    \notag\\
    &=
    \lie_{\T} *( \J^{\flat} \wedge \dd \J^{\flat} ) - \iota_{\T} \dd *( \J^{\flat} \wedge \dd \J^{\flat} )
    \notag\\
    &= 0\ ,
\end{align}
which implies that $ *( \T^{\flat} \wedge \J^{\flat} \wedge \dd \J^{\flat} ) $ is a constant. 

Note that $ (\J^{\flat} \wedge \dd \J^{\flat} ) \big|_{r=0} \in \Lambda^3(S)$, the three-form space of $S$, while $\dim S = 2$, and hence $ *( \T^{\flat} \wedge \J^{\flat} \wedge \dd \J^{\flat} ) \equiv *( \T^{\flat} \wedge \J^{\flat} \wedge \dd \J^{\flat} ) \big|_{r=0} = 0 $. Similarly,  $*( \T^{\flat} \wedge \J^{\flat} \wedge \dd \T^{\flat} ) \equiv *( \T^{\flat} \wedge \J^{\flat} \wedge \dd \T^{\flat} ) \big|_{r=0} = 0 $, and thus the Frobenius conditions hold.

As a consequence, a stationary, asymptotically flat, $I^+$-regular, analytic vacuum spacetime $(\M,g)$ with a non-rotating and degenerate horizon component $\Hh_i$, must be a stationary and axisymmetric spacetime. By the combination of Theorem \ref{thm:HKWX1} and Lemma \ref{lem:HCA}, the horizon is in fact connected, $\Hh = \Hh_i$. By 
Theorem \ref{thm:uniq-single-bh}, this spacetime must be a Kerr spacetime. However, no member of the Kerr spacetimes has a non-rotating and degenerate horizon.
Therefore, a non-rotating and degenerate horizon component cannot exist, and we complete the proof of the \textbf{Main Theorem}.  

As stated in the introduction, since the \textbf{Main Theorem} rules out the final case in the classification of stationary, asymptotically flat, $I^+$-regular, analytic vacuum spacetimes, we prove the \textbf{Corollary}, which establishes the black hole uniqueness for these analytic spacetimes.

\par\par\medskip

\begin{remark}
    For stationary, asymptotically flat, $I^+$-regular, analytic \textit{electrovacuum} spacetimes, one may use the same method as in the present paper to obtain some uniqueness results.
    When the horizon has at least one rotating component, the Hawking's rigidity theorem remains valid, and thus the spacetime is stationary and axisymmetric in the sense of Definition \ref{def:stationary and axisymmetric}. Then combining the main theorem 1.1 by Han--Khuri--Weinstein--Xiong in \cite{HKWX2} and Remark \ref{rmk:electrovacuum1} in our paper, one can obtain that the spacetime is either the Kerr--Newman spacetime (when the horizon is connected), or the Majumdar--Papapetrou spacetime (when the horizon is disconnected).
    When all of the horizon components are non-rotating and non-degenerate, Sudarsky--Wald \cite{Sudarsky-Wald1992} proved that the spacetime is static and Chru\'sciel--Tod \cite{Chrusciel-Tod2007} proved that the static electrovacuum spacetime is either the Reissner--Nordstr\"om spacetime (when the horizon is connected), or the Majumdar--Papapetrou spacetime (when the horizon is disconnected).
    
    The final case is that all of the horizon components are non-rotating, but at least one horizon component is non-rotating and degenerate. By Colling--Katona--Lucietti \cite{Lucietti-Colling-Katona}, one can still find the axisymmetric Killing field $\J$ on the horizon component. 
    If $\J$ does not vanish everywhere on the cross-section $S$, then one can still extend it to be an axisymmetric Killing field in the domain of outer communication $\D$  with a similar method in the present paper. 
    However, when $\J$ vanishes everywhere on $S$, which is equivalent to the fact that $h_a$ is an exact form, the Gauss-Bonnet theorem does not lead to a contradiction when the spacetime is electrovacuum. Therefore, it remains an open problem to classify the electrovacuum spacetimes when $h_a$ is an exact form.

\end{remark}

\bibliographystyle{plainnat}
\bibliography{main}

\bigskip
\noindent\hspace*{1em}
\begin{tabular}{@{}l@{}}
  {\scshape Department of Mathematics, Sun Yat-sen University, Guangzhou, China}\\
  \textit{Email address:} \texttt{huangysh36@mail2.sysu.edu.cn}
\end{tabular}

\medskip
\noindent\hspace*{1em}
\begin{tabular}{@{}l@{}}
  {\scshape Department of Mathematics, Sun Yat-sen University, Guangzhou, China}\\
  \textit{Email address:} \texttt{lijunbin@mail.sysu.edu.cn}
\end{tabular}

\medskip
\noindent\hspace*{1em}
\begin{tabular}{@{}l@{}}
  {\scshape Department of Mathematics, Sun Yat-sen University, Guangzhou, China}\\
  \textit{Email address:} \texttt{stszxp@mail.sysu.edu.cn}
\end{tabular}

\end{document}